\documentclass[10pt,twocolumn,letterpaper]{article}

\usepackage[pagenumbers]{cvpr} 

\usepackage[dvipsnames]{xcolor}

\definecolor{cvprblue}{rgb}{0.21,0.49,0.74}
\usepackage[pagebackref,breaklinks,colorlinks,citecolor=cvprblue]{hyperref}
\usepackage{nicematrix,enumitem,booktabs}

\def\paperID{****} 
\def\confName{CVPR}
\def\confYear{2024}

\usepackage{hyperref}
\graphicspath{{./images/}, {./images_supp/}}
\title{MiceBoneChallenge: Micro-CT public dataset and six solutions for automatic growth plate detection in micro-CT mice bone scans}

\author{Nikolay Burlutskiy\thanks{corresponding author, email nikolay.burlutskiy@astrazeneca.com} \and Marija Kekic \and Jordi de la Torre \and Philipp Plewa \and Mehdi Boroumand \and Julia Jurkowska \and Borjan Venovski \and Maria Chiara Biagi \and Yeman Brhane Hagos \and Roksana Malinowska-Traczyk \and Yibo Wang \and Jacek Zalewski \and Paula Sawczuk \qquad Karlo Pintaric \qquad Fariba Yousefi \qquad Leif Hultin\\ AstraZeneca R\&D}

\begin{document}
\maketitle

\begin{abstract}
Detecting and quantifying bone changes in micro-CT scans of rodents is a common task in preclinical drug development studies. However, this task is manual, time-consuming and subject to inter- and intra-observer variability. In 2024, Anonymous Company organized an internal challenge to develop models for automatic bone quantification. 
We prepared and annotated a high-quality dataset of 3D $\mu$CT bone scans from $83$ mice. The challenge attracted over $80$ AI scientists from around the globe who formed $23$ teams. The participants were tasked with developing a solution to identify the plane where the bone growth happens, which is essential for fully automatic segmentation of trabecular bone. As a result, six computer vision solutions were developed that can accurately identify the location of the growth plate plane. The solutions achieved the mean absolute error of $1.91\pm0.87$ planes from the ground truth on the test set, an accuracy level acceptable for practical use by a radiologist. The annotated 3D scans dataset along with the six solutions and source code, is being made public, providing researchers with opportunities to develop and benchmark their own approaches. The code, trained models, and the data will be shared (link in supplementary material).
\end{abstract}

\section{Introduction}
\label{sec:intro}

Throughout the lifespan of vertebrates, bone tissue undergoes continuous and active remodeling, regulated by a complex interplay between two essential cell types: osteoblasts, responsible for bone formation, and osteoclasts, which handle bone resorption. This dynamic equilibrium is strongly influenced by the endocrine and immune systems, creating a finely tuned process crucial to bone health~\cite{Clarke}.

Assessing bone morphology and bone mass is valuable not only for understanding and diagnosing bone diseases such as arthritis, osteoporosis, and osteoblastomas but also for evaluating potential bone safety concerns associated with new therapeutic drugs. Preclinical studies in small animals play a critical role in examining these disease mechanisms and assessing drug safety before clinical trials in humans.

Micro-computed tomography ($\mu$CT) is a widely used technique for assessing bone remodeling effects in both clinical and preclinical specimens. It enables the acquisition of high-resolution 3D images of bone microarchitecture, allowing for the quantification of key structural parameters. Furthermore, by calibrating the scanner with hydroxyapatite of known density, image intensity can be accurately correlated with bone mineral density.

The assessment involves scanning dissected bones, followed by the manual identification and quantification of bone growth regions. These techniques are widely applied in drug development. Consequently, protocols and parameters have been established for $\mu$CT analyses in mice, specifically for long bones, vertebrae, and palms in aging mice~\cite{mct_bone_mice}.

Typical endpoints of $\mu$CT analysis include trabecular bone volume fraction, trabecular number, thickness and separation, as well as cortical cross-sectional area, area fraction, and thickness~\cite{Bouxsein}. These parameters are measured within a defined volume of interest (VOI), specified based on its distance from a distinct anatomical landmark, such as the epiphyseal growth plate of the femur or the tibial metaphysis~\cite{Shim}.

Within the VOI, cortical and trabecular bone regions are identified and segmented independently. Traditionally, separating these two bone regions has been a manual task, achieved by contouring the outer trabecular perimeter. However, these manual tasks are time-consuming and susceptible to inter- and intra-observer variability. Recent advances in AI have allowed researchers to develop and automate this process, leading to more efficient analysis of bone remodeling. For instance, automated or semi-automated solutions based on dual-thresholding~\cite{Bouxsein}, atlas-based methods~\cite{Newton}, and deep learning approaches~\cite{Mahdi, HR-pQCT} have been implemented in recent years.

However, the selection of the VOI remains a manual and time-consuming task, prone to inter-operator variability. Additionally, only a limited number of preclinical bone $\mu$CT datasets are publicly available at resolutions suitable for bone remodeling analysis. In ~\cite{Rosenhain_nat1} the authors provide a full skeleton segmentation dataset of $228$ mice at a resolution of $28$ $\mu$m, while~\cite{Ranzoni} offer a dataset with $21$ scans of the proximal tibiae of wild-type mice at $5.06$ $\mu$m pixel size, and~\cite{Zenzes} has made available scans of $19$ female C57BL/6 mice lumbar vertebrae at a resolution of $2.5$ $\mu$m.

Several studies have developed AI models to quantify changes in skeletal bones. For instance, in~\cite{old_stuff06}, the authors introduced a model using active shape models and random forest regression for automated bone age assessment. These early, pre-deep learning models required extensive feature engineering and struggled to capture the complex non-linear relationships within the data. The advent of deep learning has allowed researchers to bypass such feature engineering; for example, in~\cite{old_stuff18}, the authors employed convolutional neural networks (CNNs) to predict bone age from pediatric hand radiographs. This model achieved performance levels comparable to human experts, underscoring the promise of deep learning approaches. Another similar model for bone age assessment using deep learning was developed in~\cite{PMID:27816861}. 

However, there has been limited research on the automatic quantification of bone remodeling in animals, such as rats or mice, despite its potential to significantly accelerate preclinical drug development.

Fortunately, an increasing number of public datasets, including radiology data, have become available to accelerate AI development in this field. However, compared to human datasets, there are only a few annotated animal datasets. For bone remodeling quantification, there are currently no public datasets available and no public or commercial models for automatic bone growth identification and quantification. To address this gap, we organized an internal challenge at Anonymous Company, which ultimately led to the development of six high-performing deep learning models capable of accelerating preclinical drug development.

In this work, we tackle the challenge of automating VOI selection by developing and comparing six state-of-the-art deep learning methods for growth plate detection in distal femur scans of mice. Additionally, we prepared, annotated, and publicly shared a high-quality 3D $\mu$CT dataset of mice bone scans, including annotations for trabecular bone volume and the growth plate cortical plane.

The main technical challenge when working with high-resolution 3D data is its dimensionality. While the full 3D structure can be analyzed using techniques such as 3D convolutions, this approach is often computationally infeasible. Processing 2D slices offers a more stable model, though it sacrifices information across slices. To address this, several pseudo-3D (or 2.5D) methods have been proposed, typically involving the stacking of multiple 2D slices either before applying convolutions or by stacking the outputs from convolutional layers later in the architecture~\cite{Jendeberg2021}. A review of popular 2D, 3D, and 2.5D methods can be found in~\cite{litjens2017survey, Zhang2022}.

The main contributions of this paper are:

\begin{itemize}
    \item We prepared, annotated, and publicly shared a high-quality 3D dataset of $\mu$CT mouse bone scans;
    \item We developed and compared six state-of-the-art deep learning methods for the automatic identification of the growth plate plane in the bone scans;
    \item We publicly shared all code and the models for benchmarking and further development.
\end{itemize}

\section{Motivation: why the challenge was set up}

Bone growth quantification is an integral part of the safety assessment in drug development. Clinical research organizations and pharmaceutical companies routinely run thousands of preclinical studies per year where bone quantification is involved. Each study may include hundreds of animal bones for quantification. Currently, the quantification process has many manual steps, including visual exploration of each bone, the identification of the growth plate, and then the quantification of its parameters. Finally, the domain expert needs to segment out the growth area at the growth plate and quantify the area. This process takes approximately five minutes per bone and can take several days for a whole bone quantification study. An automated process will not only eliminate the manual annotation step but will also eliminate inter- and intra-observer variability. Minimizing observer variability is important since even relatively small drug-induced changes of $\pm$ $10$\% in bone volume or bone density can have a significant impact on the progression of therapies. Finally, the automated approach can be easily scaled, allowing one to run several studies in parallel.

\begin{figure}[t]
  \centering
   \includegraphics[width=1.0\linewidth]{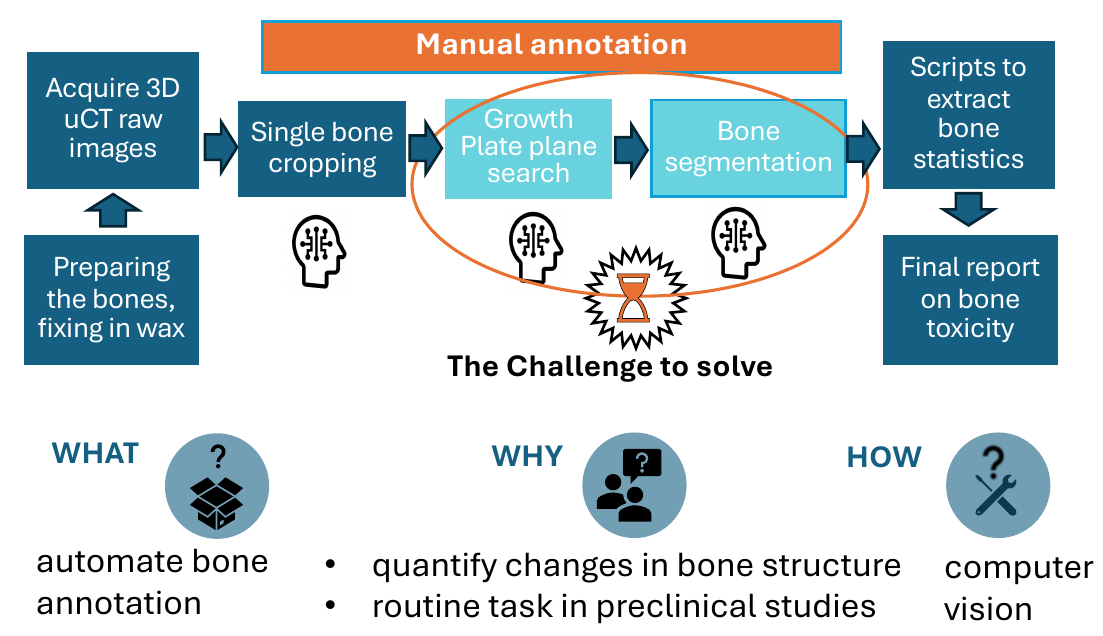}
   \caption{The process of bone quantification has several manual annotation steps that motivated us to organize the challenge. Growth plate plane search and then bone segmentation were the two tasks in the challenge. This paper is focused on automatic growth plate plane detection solutions.}
   \label{fig:problem}
\end{figure}

\section{Challenge set up}

The end-to-end process was split into two tasks for the challenge: the first was to identify the growth plate, and the second was to segment and quantify the bone at the growth plate. The second step is quite straightforward even with simple U-net like architectures, but the first task is quite challenging, so the paper covers six independent solutions that were developed by six Anonymous Company teams to solve the task.

The Challenge was aimed at data scientists of all levels of experience. We aspired to bring colleagues from different functions together, so they can get to know each other’s expertise and create opportunities for the flow of cross-functional ideas and new collaborations. The event also provided opportunities to learn and develop participant's knowledge of imaging data science. Training on our internal AWS-based platform was also provided. The challenge was launched on March 4th, with final presentations on July 2nd, 2024. Each team had the freedom to determine their own working patterns. 

\subsection{Growth plate plane (GPP) definition} 

In the bone analyses, dissected femur bones from mice are scanned at a high resolution ($10\times10\times10\mu\textrm{m}$) and reconstructed as CT volume images, with voxel intensity given in Hounsfield units (HU). The primary output of the analysis is trabecular bone density and volume, calculated in a volume of interest (VOI). This VOI is a volume of $1$ mm length, starting $0.3$ mm distal to the growth plate plane (GPP) of the femur.

\begin{figure}[t]
    \centering
    \includegraphics[width=1.0\linewidth]{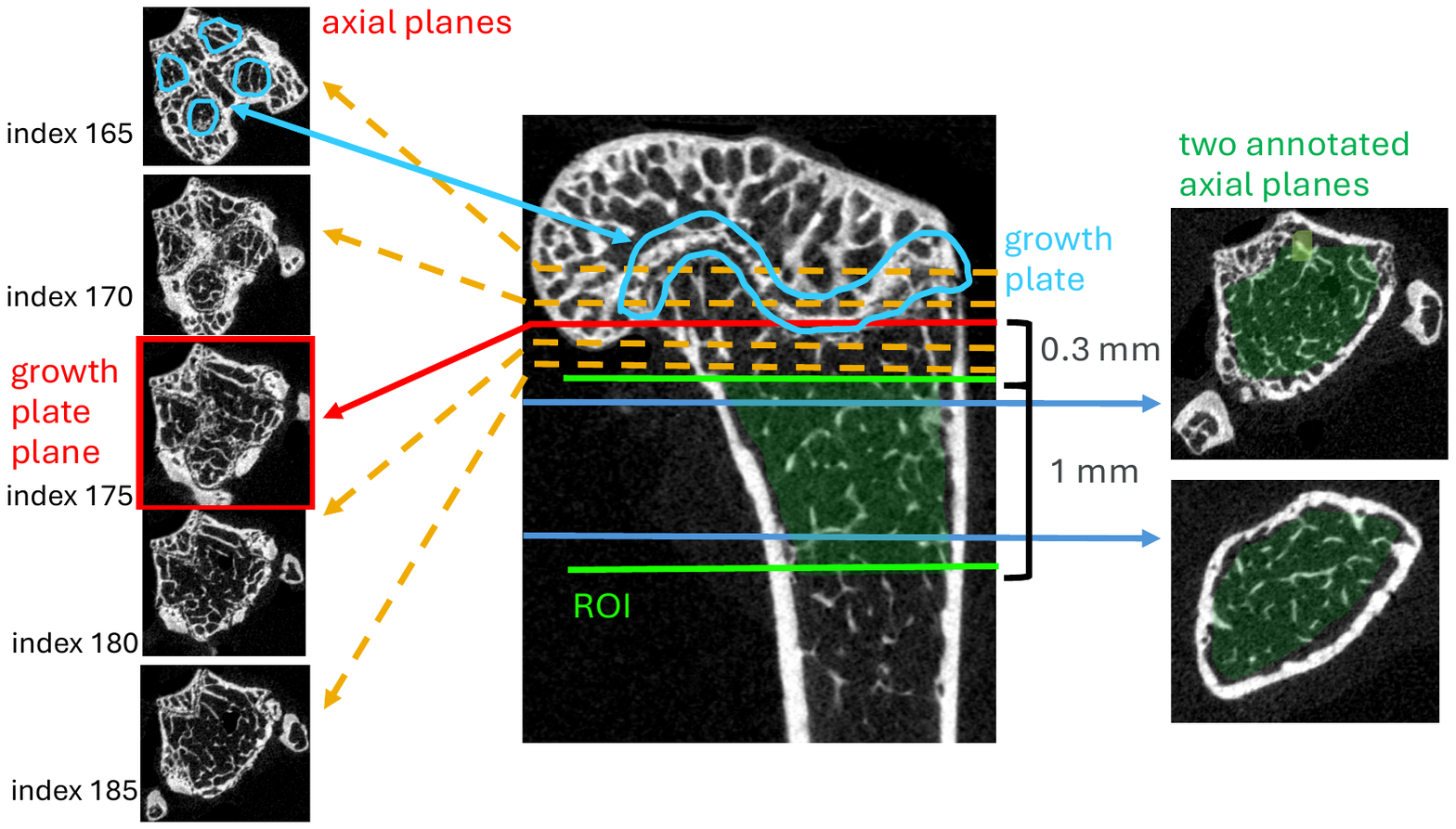}
   \caption{The challenge is to predict the growth plate plane index (GPPI). The growth plate plane is defined as the lowest plane of the growth plate (the blue area) where the bone grows. In this Figure GPPI equals $175$, it is the red plane in the axial projection.}
   \label{fig:growth_plate}
\end{figure}

The GPP that determines the VOI to quantify, is defined as the lowest plane of the growth plate. The blue areas in Figure~\ref{fig:growth_plate} depict the growth plate. The GPP is highlighted in red in Figure~\ref{fig:growth_plate}, it is the axial plane in which all four protrusions (blue oval areas in the axial plane) at the head of the femur merge (the axial slice is outlined in red, corresponding to the red arrow). 

Currently, the GPP is identified visually by the image analysts, comparing the consecutive slices along the bone main axis. Once the GPP is identified, trabecular areas (the green areas in the cyan axial plane in Figure~\ref{fig:growth_plate}) are manually outlined in the axial slices in the $1$ mm long volume, starting $0.3$ mm below the GPP, and the 3D trabecular ROI is outputted.

GPP identification and manual outlining of the trabecular regions are two operator-dependent, manual time-consuming steps. The tasks proposed in this challenge aim at finding solutions to automate both processes: Task 1 is to segment the bone (green areas in the cyan axial plane in Figure~\ref{fig:growth_plate}), Task 2 is to predict the GPPI, for example GPPI=175 for the bone in Figure~\ref{fig:growth_plate}.

\subsection{Data: $\mu$CT scanning and 3D reconstruction}

To set up the challenge, a $\mu$CT dataset was prepared, annotated, curated, and divided into training and test sets before being shared with the challenge participants.

\textbf{Data acquisition and preparation.} Mice femur were dissected, fixed in formalin for 48 hours, and then stored in 70\% ethanol. The bones, in batches of four per sample, were fixed with wax mold within a 3D printed holder and wrapped in ethanol soaked cloth before undergoing $\mu$CT scanning. Scanning was performed at a $10$ $\mu$m resolution using a SCANCO vivaCT-40 system (SCANCO Medical AG, Fabrikweg, Switzerland; 55 peak kilovoltage and 145-$\mu$A X-ray source). The 3D reconstruction of the scanned $\mu$CT images were completed using the Scanco scan software and exported as DICOM files. The resulting DICOM images, with dimensions of $2048\times2048\times642$ and 8-bit pixels format, were further cropped into per-bone volumes. Each DICOM scan produced four individual volumes, approximately sized $ 400\times 400\times642$. The prepared dataset comprised images from three preclinical studies, including $83$ mice, with one image per mouse, totaling $83$ $\mu$CT images.

\textbf{Annotations.} Once the imaging datasets was acquired, the scans were annotated by a domain expert with 20 years of experience of conducting in-vivo $\mu$CT scanning and quantifying the images for preclinical studies. The annotation process involved two steps. First, identifying the growth plate in the 3D image and then annotating trabecular bones in a corresponding VOI (see Figure~\ref{fig:growth_plate} and the previous subsection for more explanations). The annotated GPP index (GPPI) is then served as a reference plane for the challenge described in this paper. Finally, a quality control of the annotations and the data were performed by another domain expert.

\textbf{Training and test split.} The annotated dataset was split into training and test sets, stratified by the preclinical studies. The training set had $70$ 3D images and the test set had 13 3D images. The training set was shared with the teams for model development and then the test set was used to calculate performance metrics for the developed models and finally to rank the teams and to determine the winners.

\subsection{Challenge logistics and evaluation}

The challenge attracted more than $80$ AI scientists and engineers who formed $23$ teams. The teams received the problem statement, the training data as well as got connected to domain experts who had provided guidance regarding bone anatomy, data acquisition process, as well as guidance for computer vision algorithms and deep learning tools. Then after $4$ months the teams received the test dataset and were asked to provide predictions of the GPPI for each of $13$ 3D bone scans. Finally, six teams provided solutions for the challenge of automatic GPPI detection. Then the submitted models were used to run inference on the test set followed by calculating a mean score, standard deviation, and the sum of scores for each team. The score for each bone was calculated using an evaluation function (more details in subsection~\ref{EvaluationFunction}) to penalize GPPI predictions far from the ground truth. Then the teams were ranked based on the sum of the scores and the top $3$ teams were awarded with prizes.

\subsubsection{Evaluation Function}\label{EvaluationFunction}

To assess model performance, we employed a scaled survival function based on the standard normal distribution. The evaluation function is defined as:

\begin{equation}
    \text{Score} = 2 \times \bar{\Phi}\left(\frac{p - t}{3}\right)
    \label{eq:score_formula}
\end{equation}

where $p$ is the predicted plane number, $t$ is the true (tagged) plane number, and $\bar{\Phi}$ is the complementary cumulative distribution function (CCDF) of the standard normal distribution. This function provides a smooth decay in the score as the difference between the predicted value $p$ and the true value $t$ increases, with a scaling factor of $3$ in the denominator to control the rate of decay. The multiplication by $2$ scales the output to a range of [0, 1], where 1 represents a perfect prediction (difference of $0$), and the score decreases as the difference increases. This approach allows for some tolerance in predictions while still penalizing larger errors, providing a balanced assessment of our model's accuracy in predicting GPPs.

Table \ref{tab:score_distance} illustrates the relationship between the predicted plane error $e$ defined as the difference of predicted $p$ and true plane $t$ numbers and the resulting score.

\begin{table}[ht]
\centering
\caption{Predicted plane error $e$ vs evaluation score}
\label{tab:score_distance}
\resizebox{0.47\textwidth}{!}{
\begin{tabular}{ccccccccc}
\hline
$e = |p - t|$ & 0 & 1 & 2 & 3 & 4 & 5 & 6 & 7 \\
\hline
\textbf{Score} & 1.00 & 0.74 & 0.50 & 0.32 & 0.18 & 0.10 & 0.05 & 0.02 \\
\hline
\end{tabular}
}
\end{table}

As shown in Table~\ref{tab:score_distance}, the score decreases rapidly as the difference between predicted and true plane numbers increases, emphasizing the importance of accurate predictions in our model.

\subsection{Challenge restrictions}

For the challenge, an AWS environment with access to GPU enabled instances was set up for the participants. The organizers prepared pre-processing and evaluation scripts to facilitate a quick onboarding for the challenge. The participants had the freedom to choose any model architecture, provided that the training would be possible to run on $1$ GPU NVIDIA T4 with 16GB of memory. Ensembling the models was allowed with no restrictions on the number of the models in the ensemble. The participants were allowed to use any pre-trained models if they could find any in the public domain, same freedom was provided for using any external datasets. Communication and knowledge sharing between the teams were allowed so the teams could share ideas and code, for example, code for pre-processing that facilitated collaboration. No restrictions on software, frameworks, programming language, software libraries were imposed. The organizers reserved the right to disqualify the participants if the participants did not follow the challenge’s rules.

\section{Submitted Solutions}

\begin{figure}[t]
  \centering
   \includegraphics[width=1.0\linewidth]{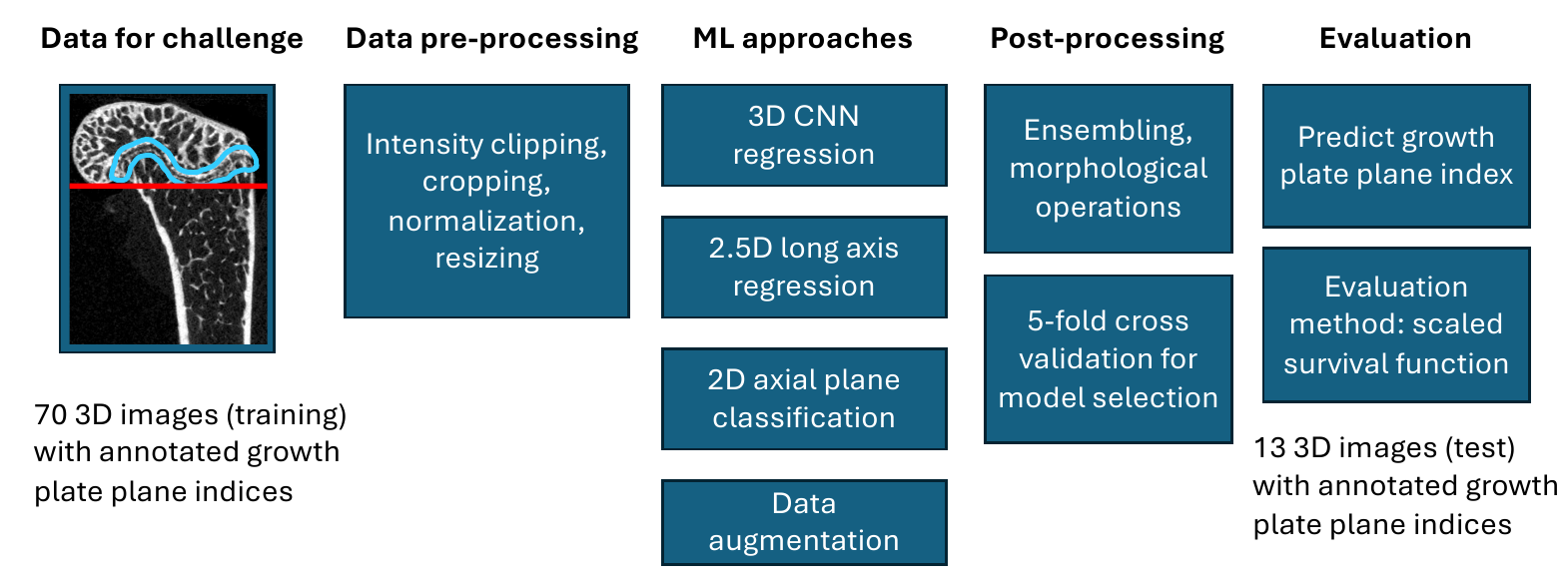}
   \caption{Six teams applied different pre-processing techniques, utilized 3D/2.5D/2D modeling approaches, used data augmentation, and then utilzed post processing methods including ensembling and cross validation. Finally, the performance of the teams was evaluated using a survival function for ranking the teams.}
   \label{fig:overview}
\end{figure}

An overview of the six solutions from six teams submitted for evaluation, is summarized in Figure~\ref{fig:overview}. The teams were \textbf{SafetyNNet}, \textbf{Matterhorn}, \textbf{CodeWarriors2}, \textbf{Exploding Kittens}, \textbf{Subvisible}, and \textbf{ByteMeIfYouCan}. Later in the paper these teams are referred as Teams \textbf{SN}, \textbf{MH}, \textbf{CW}, \textbf{EK}, \textbf{SV}, and \textbf{BM} accordingly. 

Next, we outline data pre-processing, modeling approaches, data augmentation, and finally cross validation and ensembling approaches the teams employed in their solutions. More details for each approach are in Supplementary material.

\subsection{Data Pre-processing} In the data pre-processing phase, images underwent clipping, normalization, cropping of input voxel values, and resizing to enable batching of the images. Several teams also scaled the images. Finally, for modeling, the teams used 3D, 2.5D and 2D images. Team SN used 3D voxels as input for the modeling, MH and EK used 2.5D images - 2D images enriched with partial information from the longest axis. CW, SV, and BM used 2D input using an independence assumption for the 2D images across the longest axis. The pre-processing details for each team are in Table \ref{tab:data_preprocessing}.

\begin{itemize}
\item The majority of teams applied clipping to HU values using ranges from $-1000$ to $500$ for the low range and from $1900$ to $4000$ for the high range. This clipping of HU scale reduced image noise, focusing specifically on bone density. 
\item Due to variations in the axial (xy) dimensions of the images, all the teams utilized cropping and then interpolation techniques to resize the images to the desired dimensions, depending on the deep learning network backbone used. 
\end{itemize}

\begin{table}[ht]
\centering
\caption{Pre-processing steps and parameters for each team}
\resizebox{0.47\textwidth}{!}{
\begin{tabular}{@{}cccccc@{}}
\toprule
Team & Clipping & Cropping & Resize & Interpolation & Other \\ \midrule
SN & $-1000, 4000$ & $448\times448$ & $321\times244\times244$ & Linear & 3D \\
MH & $-1000, 3000$ & $300\times300$ & No           & Linear & 2.5D, Scaling \\
EK & $500, 2000$   & $515\times515$ & No           & No & 2.5D, Scaling\\
CW & $0, 1900$     & $322\times307$ & No           & No & No\\
SV & No            & No           & $96\times96\times1$   & Linear & Scaling \\
BM & $-100, 3171$  & $480\times480$ & $240\times240\times1$ & Area  & Scaling \\
\bottomrule
\end{tabular}
}
\label{tab:data_preprocessing}
\end{table}

\subsection{Modeling Approaches}

All six solutions can be divided into three groups, each group representing the approach chosen, namely i) 3D CNN regression approach utlizing full 3D volume data (Team SN), ii) 2.5D long axis regression approach from an appropriate 2D sagittal or coronal view (Teams MH, EK), and finally iii) 2D/2.5D axial plane classification approach (Teams CW, SV, BM).

All 2.5D approaches used 3rd dimension as channels, effectively applying 2D convolutions across axial plane and fully connected layer across the 3rd dimension.

\subsubsection{3D CNN Regression Approach} The utilization of full 3D information provides the most comprehensive data for the model to derive insights. However, due to high dimensionality, only crops of the full volume could be processed on the GPU.

Team \textbf{SN} opted for a 3D convolutions using a sliding 3D window approach along the long axis of the bone. They used ResNet34 for feature extraction and a decoupled head, similar to object detection pipelines \cite{ge2021yoloxexceedingyoloseries}. The two outputs of the two heads were a probability of whether the GPPI was contained in the crop and a regression offset of the GPPI from the beginning of the crop, scaled to the crop length. The loss function was a weighted sum of the two terms, comprising Sigmoid Focal Loss for classification and Mean Squared Error for regression, with the second term being masked in cases where the GPPI was not within the crop. During prediction, a window with the maximal classification prediction was selected, and the exact GPPI was calculated using the regression offset prediction.

The performance across the five folds for SN and three other teams who used cross validation for model selection is summarized in Table~\ref{tab:cross_validation}.

\subsubsection{2.5D Long Axis Regression} 

GPPI can be inferred from an appropriate 2D view of a long axis (sagittal or coronal). To ensure the correct long axis view is contained in the input images, one can utilize stacking of various slices across the channel dimension. Two teams, \textbf{MH} and \textbf{EK}, utilzed such stacking approach. The final output of the network is a regression coefficient offset of GPPI from the image beginning, scaled to the image length to produce a label in the $0-1$ range.

Team \textbf{MH} used sorted random slices from the sagittal plane, chosen from an internal portion of the dimension. This approach maximizes the probability of selecting locations containing useful bone information and avoids slices that contain only a small portion of the bone volume. The backbone was based on EfficientNet B3 with modified input channel dimensions. Additionally, the team applied a dual training process, passing images in both low resolution (containing the full length of the bone) and high resolution (containing only the crop around the GPPI). The inference process then consists of two evaluations: an initial interpolated prediction to broadly identify the growth plane’s location, followed by a more precise cropped prediction within the boundaries established in the first step.

In Table \ref{tab:cross_validation}, we present the results of various architectures evaluated through a comprehensive five-fold cross-validation process. This table summarizes the performance metrics, including mean scores and standard deviations, for each model across all folds. 

\begin{table}[ht]
    \centering
    \caption{Five-Fold cross validation metrics for four teams}
    \resizebox{0.47\textwidth}{!}{
    \begin{tabular}{@{}lcccccccc@{}}
        \toprule
        Model & $\mu_{score}$ & $\sigma_{score}$ & Fold 1 & Fold 2 & Fold 3 & Fold 4 & Fold 5 \\ \midrule
        Team SN &&&&&&& \\
        \hline
        ResNet34  & 0.605   & 0.050  & 0.632  & 0.626  & 0.664  & 0.518  & 0.585  \\ 
        \hline
        Team MH \\
        \hline
        B0-300  & 0.578   & 0.017  & 0.605  & 0.577  & 0.580  & 0.559  & 0.570  \\
        B1-300  & 0.604   & 0.053  & 0.667  & 0.616  & 0.585  & 0.626  & 0.526  \\
        B2-300  & 0.587   & 0.022  & 0.573  & 0.560  & 0.611  & 0.581  & 0.609  \\
        B3-300  & 0.580   & 0.020  & 0.591  & 0.566  & 0.602  & 0.553  & 0.586  \\
        B4-300  & 0.590   & 0.033  & 0.629  & 0.569  & 0.564  & 0.622  & 0.565  \\
        B4-380  & 0.585   & 0.046  & 0.589  & 0.623  & 0.514  & 0.574  & 0.627  \\ 
        \hline
        Team SV &&&&&&& \\
        \hline
        Custom  & 0.542 & 0.055  & 0.584  & 0.588  & 0.438  & 0.536  & 0.566  \\ 
        \hline
        Team BM &&&&&&& \\
        \hline
        DenseNet  & 0.548 & 0.107  & 0.553  & 0.525  & 0.739  & 0.410  & 0.515  \\ 
        \bottomrule
    \end{tabular}
    }
    \label{tab:cross_validation}
\end{table}

The model selected for final test evaluation was B3 with a size of 300. This decision was made after careful consideration of the performance metrics across the various models. Notably, there were no significant differences in the mean scores among the models evaluated, which included B0, B1, B2, B3, and B4 with a size of $300$. Given this lack of substantial variation in performance, we opted for the standard B3-300 model as a reliable benchmark for our evaluation.

Team \textbf{EK} constructed 3-channel 2D images by stacking slices from sagittal, coronal, and a blend of the two slices. From each bone scan, $14$ different sagittal and coronal slices were taken, starting at the center $(x_0, y_0)$ and offsetting further slices by $\pm5$, resulting in a total of $49$ images per 3D scan. The backbone was a pretrained EfficientNet-es pruned.

\subsubsection{2D/2.5D Axial Plane Classification} 

In principle, binary classification across the axial plane can be used to classify GPPI versus other planes. However, subsequent planes would be too similar, making naive classification ineffective.

Two teams, \textbf{CW}, \textbf{SV}, solved the challenge by redefining the classification task, taking notice that slices before and after the GPPI are distinguishable. They redefined the classification into pre-GPPI and post-GPPI. The main challenge with this approach was to denoise the possibly inconsistent output of the classifier, especially for adjacent to the GPPI planes. 

Team \textbf{CW}, utilized morphological closing filters with kernel size $5\times1$ to denoise the sequence of outputs from the binary classifier, thus making the distinction between the 'before' and 'after' classes explicit. The last slice predicted as 'before' was used as the prediction of GPPI. The CNN backbone was a pre-trained ResNet18~\cite{he2016deep} with an adapted architecture to accept grayscale input and output binary classification.

Team \textbf{SV}, opted for a two-stage training approach using two lightweight CNN networks, one for classification and one for regression, each with less than 40k total parameters. The solution is based on the observation that axial plane images display four distinct blobs, each corresponding to a protrusion, prior to the GPP, and these blobs merge and disappear after the GPP. A binary classifier was trained to identify images with these four blobs, marking the last image in the positive series as a rough estimate of the GPP. To refine these estimates, the team leveraged cross plane information, employing a regression CNN to examine 25 slices before and after the initial estimate, forming a stack of 51 images. This stack was used to generate more accurate predictions of the GPP, combining initial binary classification with regression-based refinement for better precision. These two low-parameter models combined with reducing the input size to $96\times96$ keep the computational cost of this approach low and make it very fast to train. Table~\ref{tab:cross_validation} shows the performance of this approach on the five-fold cross-validation. The mean and standard deviation for $10$ runs (training from scratch) on a single fold, specifically fold $5$, were $0.517$ and $0.036$, respectively.

Finally, Team \textbf{BM} used sliding window approach similar to Team \textbf{SN} but instead of full 3D convolution utilized stacking across channels and applying 2D convolutions.
Team \textbf{BM} passed each 3D crop through a custom DenseNet backbone to predict a value $P=0...1$, where $P=0$ indicated that the GPPI was not contained in the window, $P=1$ indicated that it was located at the center, and $P$ varied linearly with the distance from the GPPI. Binary Cross Entropy Loss was used to minimize the difference between the predicted and true $P$ target. During inference, a sliding window approach was used, where the window with the maximum value of $P$ was selected as the one with the centered GPPI. 

\subsection{Data Augmentation}

The standard data augmentation techniques included random flipping, blurring, noise addition, motion simulation, spike introduction, and ghost artifact generation. Additionally, Team \textbf{MH} implemented axial plane displacement augmentation to ensure a uniform distribution of the GPPI across the image when passing full-resolution images through the network.

\subsection{Cross validation and ensembling}

Due to the limited amount of data, four teams including SN, MH, SV, and BM employed five-fold cross-validation, see Table~\ref{tab:cross_validation} for detailed results. Also, all teams except CW applied ensembling strategy for their final predictions. The teams used ensambles of up to five models to run inferences and then averaged the predictions, more details  on the final models used for the inference are in Table~\ref{tab:model_complex}.

\subsection{Code, Models, Data and Annotations}

All the code and the models for each team as well as detailed instructions how the models were trained and how to run inference are in the supplementary material.


We share the data with the annotated GPPIs and the pixel wise segmented bone annotations. A link to the data is in the supplementary material, along with a link to the code and the trained models. 

\section{Results and Evaluation}

For evaluation, we utilized both qualitative and quantitative measurements. The qualitative assessment included comparing the predicted GPPIs and True GPPI visually (see Figure~\ref{fig:bone_gt_examples})  whereas the quantitative measurements included the differences in the number of planes between the predicted GPPI and the True GPPI (see Table~\ref{tab:final_results}), as well as the scores using the survival function (see Table~\ref{tab:performance}). We also compared the approaches in terms of complexity, number of parameters, and computational cost (see Table~\ref{tab:model_complex}). 

\subsection{Test Set Evaluation}

First, each team ran inference on the test set of 13 3D $\mu$CT bone scans and predicted the GPPI for each bone in the test set. The differences between the predicted GPPIs and the ground truth for each bone of the test set by the teams are presented in Table~\ref{tab:final_results}. The positive numbers in the table indicates that a team predicted the plane above the GPP and below the GPP for the negative numbers. The teams achieved the mean absolute error (MAE) ranging from $3.62$ for EK down to $1.23$ for BM with the mean MAE of $1.91\pm0.87$ that is acceptable performance for practical use by a radiologist.

For ranking the teams in the challenge, the evaluation function~\ref{eq:score_formula} was applied to penalize predictions far from the ground truth. Then the mean score and the standard deviation for the whole test set were calculated. The test set performance metrics include the mean score, the standard deviation of the score (STD score), and the sum of the scores that is $1$ for a perfect prediction per bone and has a maximum of $13$ for the whole test set of $13$ bones. The sum of the scores was used for ranking the teams and determine the winner. Information on the final results, including the mean scores, standard deviation, sum of the scores are in Table~\ref{tab:performance} and information on the complexity of the models, their ensembles, the number of inferences and the number of model parameters are in Table~\ref{tab:model_complex}. 

\begin{figure}[t]
  \centering
   \includegraphics[width=1.0\linewidth]{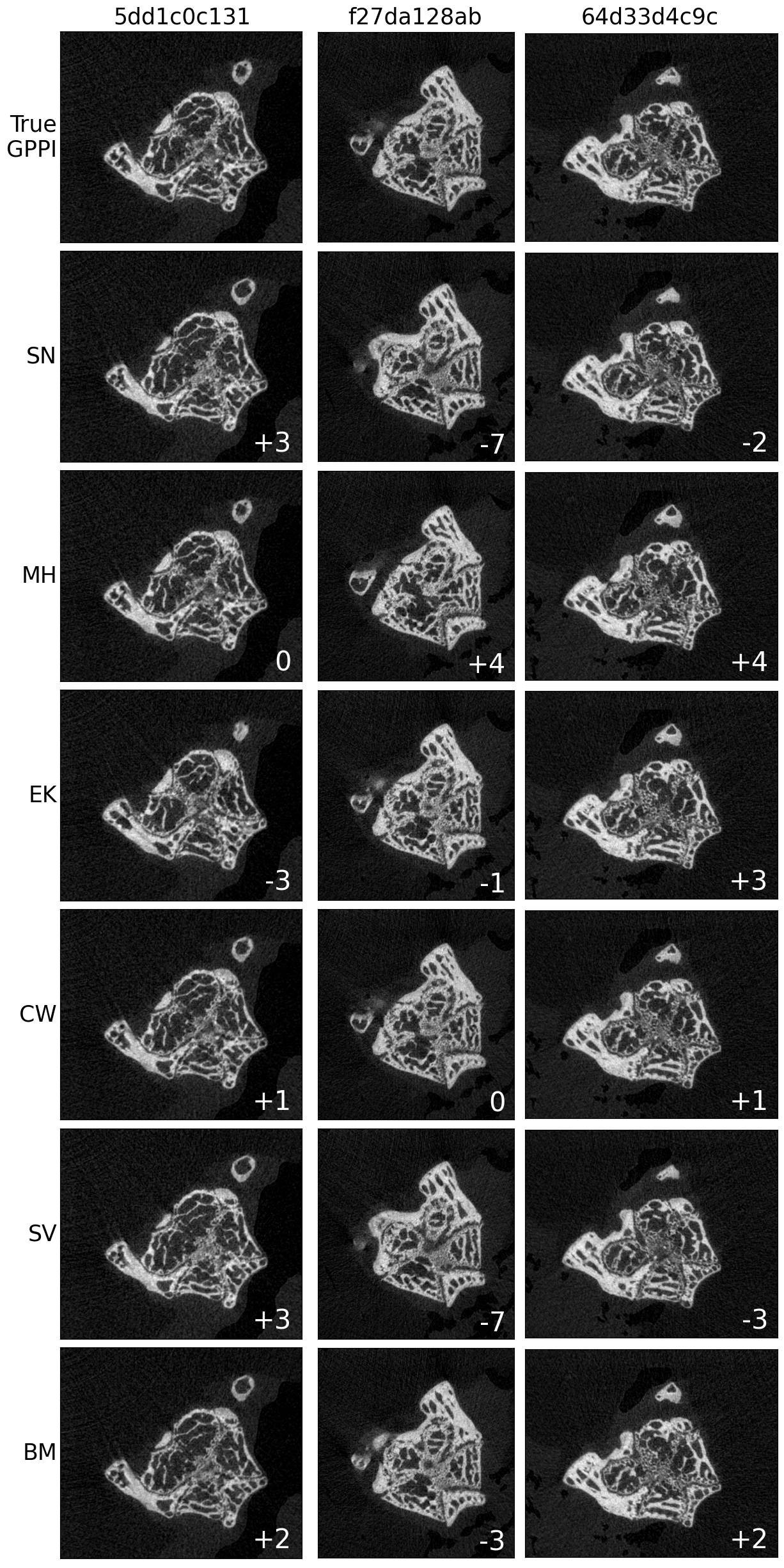}
   \caption{The examples of the True GPPI and the predictions by all the six teams for three bones, 5dd1c0c131, f27da128ab and 64d33d4c9c. For example, BM predicted GPPI +2 from the True GPPI for the bone 5dd1c0c131. The differences between the True GPPI and the adjacent growth plate planes are subtle and distinguishable only by experienced radiologists. }
   \label{fig:bone_gt_examples}
\end{figure}

Examples of the True GPPI and the predictions by all the six teams for three bones, 5dd1c0c131, f27da128ab and 64d33d4c9c are in Figure~\ref{fig:bone_gt_examples}. One can notice that the differences between the True GPPI and the predicted GPPIs that are adjacent growth plate planes are subtle and distinguishable only by experienced radiologists.

\begin{table}[ht]
\centering
\caption{Overview of the final models used for the inference}
\label{tab:model_complex}
\resizebox{0.47\textwidth}{!}{
\begin{NiceTabular}{@{}llrccc@{}}
\toprule
Team & Model & Params & Input Size & Ensembles & Inferences \tabularnote{Inferences per ensemble assuming batch of 1}\\ 
\midrule

SN  & 3D Resnet34           & ~63M  & $244\times244\times32$ & 5 & 18\\ 
\hline
MH  & EfficientNet B3       & 12.2M & $300\times300\times9$    & 5 & 10\\ 
\hline
EK  & EfficientNet-es pruned & 4.2M & $515\times515\times3$ & 4 & 4\\ 
\hline
CW  & ResNet18              & ~11M  & $322\times307\times1$ & 1  & 642\\ 
\hline
SV  & Classification CNN    & 0.04M & $96\times96\times1$ & 1 & 642\\
        & Regression CNN    & 0.04M & $96\times96\times51$ & 5  & 1\\ 
\hline
BM  & DenseNet              & ~1.4M & $224\times224\times64$ & 5 & 579\\
\bottomrule
\end{NiceTabular}
}
\end{table}

\begin{table}[ht]
\centering
\caption{Difference in the number of planes from the ground truth for the predicted GPPIs per bone from the test set and the mean absolute error (MAE) per each of six teams}
\label{tab:final_results}
\resizebox{0.47\textwidth}{!}{
\begin{NiceTabular}{@{}cccccccc@{}}
\toprule
 &  & \multicolumn{6}{c}{Team} \\ \cmidrule(lr){3-8}
Image name & True GPPI & SN & MH & EK & CW & SV & BM \\ \midrule
\hline
feaec917f3 & 199 &  0 &  0 & -2 & +1 & -3 &  0\\
a82c3c2965 & 165 &  0 & -2 & -4 & -4 & +1 & +1\\
8c9b119aa2 & 185 & -1 &  0 & -2 & -1 & +1 & -2\\
5dd1c0c131 & 180 & +3 &  0 & -3 & +1 & +3 & +2\\
f27da128ab & 175 & -7 & +4 & -1 &  0 & -7 & -3\\
6eb0a13a01 & 169 & -2 & -1 & -7 & -2 & +1 & -2\\
eda8dfa0ed & 173 &  0 & +1 & -4 &  0 & +2 &  0\\
faf967f332 & 180 & -1 & -6 & -10 & +2 & -2 & -2\\
d7cd090ca8 & 190 & +2 & -1 & +4 & +3 & -1 & -1\\
64d33d4c9c & 182 & -2 & +4 & +3 & +1 & -3 & +2\\
9bc08e060d & 164 &  0 &  0 & +2 & +3 &  0 &  0\\
7081e041aa & 199 &  0 &  0 & +4 & -1 & +1 & -1\\
22e989cc9b & 160 & +1 & -1 & -1 & -3 &  0 &  0\\
\hline
MAE & - & 1.46 & 1.54 & 3.62 & 1.69 & 1.92 & \textbf{1.23}\\
\bottomrule
\end{NiceTabular}
}
\end{table}

\begin{table}[ht]
\centering
\caption{Performance metrics for each team after applying the survival function~\ref{eq:score_formula} used for ranking the teams in the challenge}
\label{tab:performance}
\resizebox{0.47\textwidth}{!}{
\begin{NiceTabular}{@{}ccccccc@{}}
\toprule
 &  & \multicolumn{6}{c}{Team} \\ \cmidrule(lr){2-8}
Performance metrics & SN & MH & EK & CW & SV & BM \\ \midrule
\hline
Mean score &  \textbf{0.697} & 0.682 & 0.337 & 0.603 & 0.590 & \textbf{0.697}\\
STD score &  0.301 & 0.334 & 0.235 & 0.255 & 0.279 & 0.242\\
Sum of scores (max 13) &  \textbf{9.068} & 8.870 & 4.377 & 7.839 & 7.676 & 9.059\\
\bottomrule
\end{NiceTabular}
}
\end{table}

\section{Discussions}

The results from each team were sufficiently accurate for practical use by Anonymous Company's internal domain experts. However, to establish precise acceptable error rates, study-dependent endpoints obtained from both manual and automatic region-of-interest (ROI) segmentations need to be compared. Therefore, until the model demonstrates a high enough level of trustworthiness to be fully automated, a user experience (UX) design should incorporate an efficient expert-in-the-loop system, enabling experts to conduct quality control and make corrections when necessary. Implementing such a system, for instance, within Slicer3D software~\footnote{https://www.slicer.org/}, would require not only adherence to acceptable performance metrics but also reasonable inference time per image and compatibility with CPU-only machines, to minimize data transfer delays. Utilizing full 3D convolutions (Team SN) did not yield accuracy improvements over the computationally simpler approach of stacking across axial plane and applying 2D convolutions (Teams CW, SV, and BM). Among the proposed methods, the long-axis regression approaches, which require only one to two inference passes per image, proved to be the fastest solutions. However, these approaches rely on selecting the 'correct' view along the long axis, which may limit their accuracy. Potentially, a hybrid approach of combining the long-axis regression approach for the rough GPPI region with the 2.5D approaches across axial plane for finer estimation, could offer the optimal balance of speed and accuracy.

While this paper primarily addresses the challenge of identifying the GPPI, other studies have highlighted additional characteristics of the GPP itself, such as the width of the growth plate~\cite{Mango2004}, which can only be obtained by segmenting the the GPP region. Additionally, this segmentation could offer valuable insights into instances of incorrectly predicted GPPIs. Future research will explore the feasibility of extracting such segmentation details from GPPI prompts, potentially enhancing model performance and interpretability.

\section{Conclusions}

We prepared a comprehensive, high-quality 3D $\mu$CT imaging dataset with $83$ mouse bones, including growth plate plane annotations, and thorough quality curation of the dataset. Following this, we organized an internal Anonymous Company challenge that led to the development of six unique solutions utilizing different data pre-processing methods, 3D CNN regression, 2.5D long axis regression, and 2D axial plane classification modeling approaches. The proposed solutions achieved performance acceptable for practice use by radiologists. 

We are now sharing this valuable dataset along with its annotations, the developed models, and the code. 
One of the primary challenges in medical imaging lies in the substantial volume of data, rendering many existing models, which are trained on natural images, unsuitable for use in medical applications. Our aim is that, by providing this dataset and a range of solutions, more specialized models and pipelines can be developed to accelerate pre-clinical drug development. 

Recently, there have been large interest in building foundation AI models that require diverse high quality datasets like we are sharing. Ultimately, these AI models have the potential to enable drug developers to bring new medicines to patients more quickly, contributing to a significant positive impact on patient care and outcomes as well as minimize animal use in the research contributing to animal well-being.

\section{Acknowledgments}

Many thanks to The Anonymous Company for sponsoring the challenge, the efforts in acquiring and annotating the data, and all the participants.

{\small\bibliographystyle{ieeenat_fullname}
    \bibliography{ref}
}

\clearpage
\setcounter{page}{1}
\maketitlesupplementary

\section{Code, data, and models are shared}

\label{sec:supplimentary}

The source code, annotated training and test data, and models are shared for review at \href{https://osf.io/}{anonymized link}. If the paper is accepted then the code, models for all six teams, and all the $83$ annotated 3D $\mu$CT images will be publicly shared.

\section{Additional details on approaches}

More details on the approaches chosen by six teams in the challenge including training and inference parameters.

\subsection{Team SN: SafetyNNet}

Input images are first downsampled and padded to a size of $321\times244\times244$. From these, a crop of size $32\times244\times244$ is selected. Pixel values are clipped within the range $[-1000, 4000]$ and then normalized to a $0-1$ range. These crops are then passed through a 3D ResNet34 backbone, initialized with pretrained weights from~\cite{Chen2019Med3DTL}. The resulting 3D features are processed through a MaxPool layer across the \( H \times W \) dimensions and then flattened across the \( D \) dimension to obtain a feature vector of size $2048$. A dropout layer is applied after flattening to prevent overfitting.

A decoupled head, consisting of two fully connected (FC) layers, is utilized to predict objectiveness and the growth plate plane (GPP) offset from the start of the crop. The loss function is constructed as a sum of:

\begin{itemize}
    \item \textbf{Sigmoid Focal Loss:} Applied where the target is positive only if the GPP is contained within the crop.
    \item \textbf{Mean Squared Error (MSE):} Computed between the true growth plate plane index (GPPI) offset from the start of the window and the predicted offset (after passing the regression head prediction through a sigmoid function to squash it into the $0-1$ range). This term is included only if the GP index lies within the selected window. To ensure effective learning of both tasks, the MSE loss is multiplied by a factor $\lambda$, which we set to $6$. 
\end{itemize}

\begin{figure}
    \centering\includegraphics[width=0.9\linewidth]{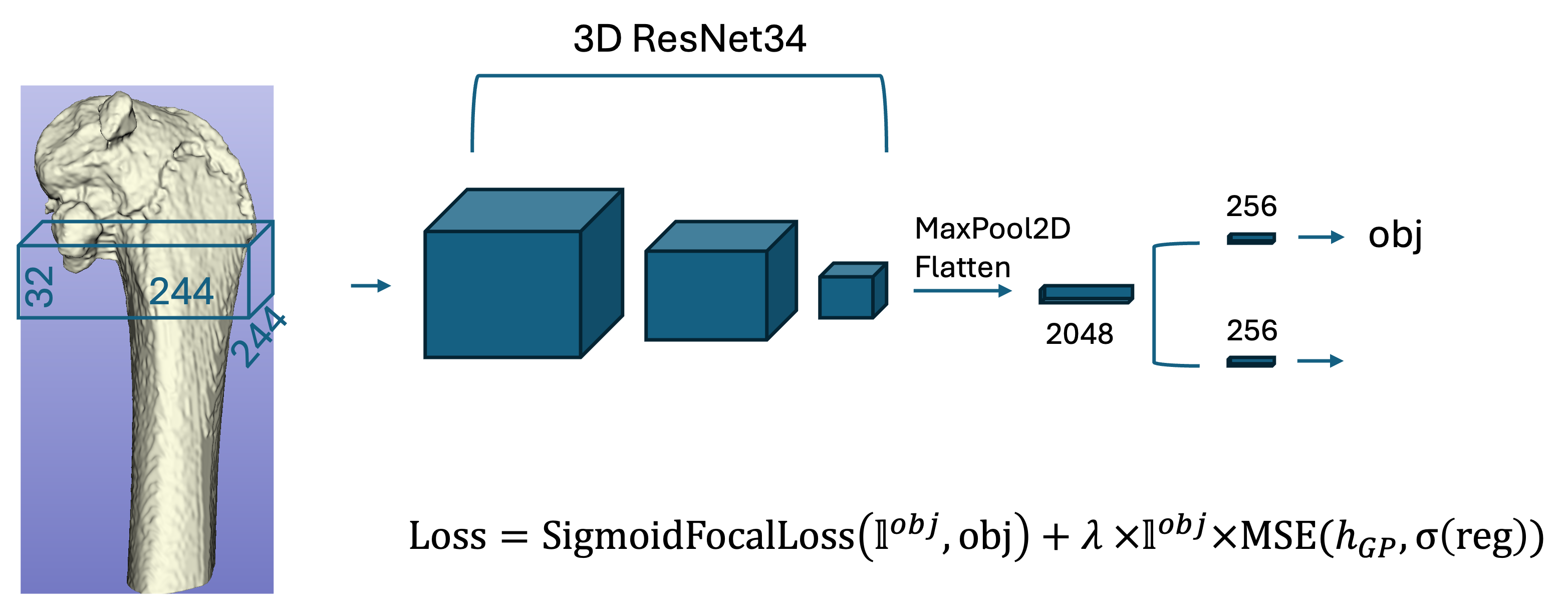}
    \caption{Training procedure by SN team utilizing 3D sliding window object detection approach.}\label{fig:Network3DSlidingWindow}
\end{figure}

The overview of the training procedure applied by SN team is given in Figure \ref{fig:Network3DSlidingWindow}. During training, crops are randomly selected to ensure balanced batches containing samples with and without growth plates. Images are augmented using the torchio library~\cite{PEREZGARCIA2021106236} with random flipping, blur, noise, motion, spikes, and ghost artifacts. Adam optimizer with exponentially decaying learning rate ($0.001$ with decay of $\gamma=0.98)$  and $10$ warm-up epochs is used. Early stopping after $100$ epoch without improvement in validation performance is applied. Training is repeated five times across different cross-validation folds, with an average score of 0.6, as explained in the main paper. 

During inference, a sliding window approach is employed over the full length of the bone, using the stride of $16$ for the donwsampled image size, and GPPI is calculated from the window with the highest classification score using the regression network offset prediction.
Final GPPI prediction is the rounded mean GPPI of five models obtained from each cross-validation split.

\subsection{Team MH:  Matterhorn}

\begin{figure}[ht]
  \centering\includegraphics[width=0.47\textwidth]{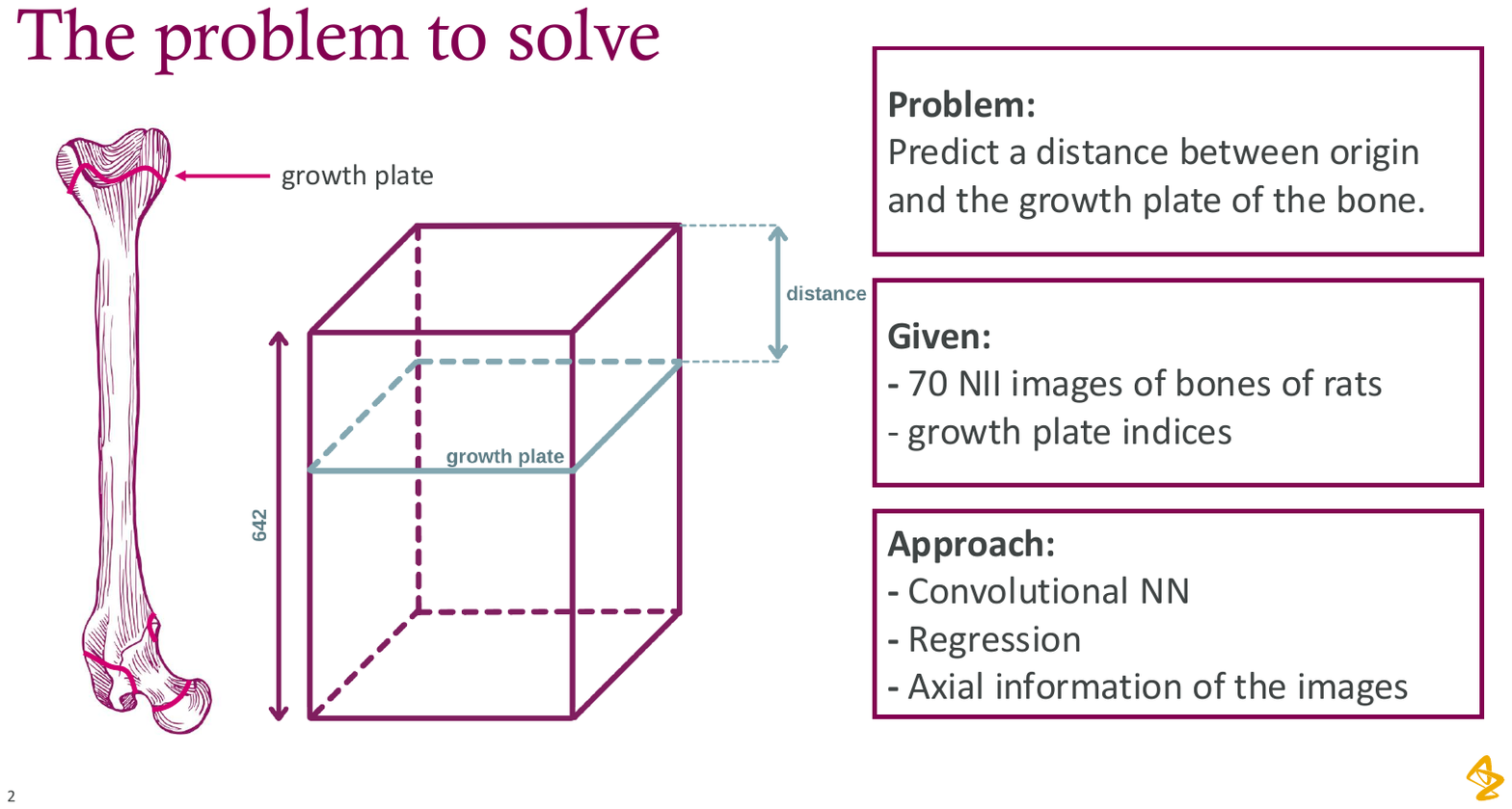}
  \caption{Predicting the growth plane plane index (GPPI) of a femur from $\mu$CT 3D images by MH team.}\label{fig:problem_matterhorn}
\end{figure}

We designed a convolutional neural network (CNN) architecture~\cite{lecun1998gradient} based on EfficientNet ~\cite{DBLP:journals/corr/abs-1905-11946} to predict the GPPI of mice femur bones using $\mu$CT scans \cite{lauterbur1973image} (see Figure~\ref{fig:problem_matterhorn}).

As explained in the main paper, we employed five-fold cross-validation to select the network architecture, as well as to provide the final ensemble of models

The cross-validation helped us to chooseEfficientNet B3, a model with the highest score and minimal variance, retaining the network's recommended input size of $300\times300$ pixels. We adapted the network to utilize $9$ input channels, selecting from sorted random axial planes of the $\mu$CT scans during both training and inference. An optimal number of channels was determined through hyperparameter tuning. Nine channels surpassed the performance achieved by configurations with less channels. Significant performance improvements were not obtained with number of channels greater than nine.

Importantly, these sorted random slices used by the network were chosen not from the entire width, but from an internal portion of the dimension. This approach maximizes the probability of selecting locations containing useful bone information beneficial for prediction, while avoiding the external limits where one might encounter the outer part of the bone or areas outside the bone entirely—neither of which is desirable for accurate predictions. The percentage of this internal portion was also optimized through hyperparameter tuning.

To strike a balance between input size constraints and prediction accuracy, we trained the network concurrently on both interpolated and cropped versions of the images. This dual approach has the advantage of requiring only one network to handle both scenarios, simplifying the model deployment. Additionally, this method introduces another source of data augmentation, which can be beneficial for improving model generalization. The interpolated case yields predictions across the entire domain with lower precision, while the cropped case provides higher precision within a confined area. This methodology shapes the inference process, which consists of two evaluations: an initial interpolated prediction to broadly identify the GPPI, followed by a more precise cropped prediction within the boundaries established in the first step.

In the pre-processing phase, we normalized the Hounsfield units (HU)~\cite{hounsfield1973computerized} to a $0-1$ range after clipping the original values between $-1000$ and $3000$.

\begin{figure}[ht]
  \centering\includegraphics[width=0.47\textwidth]{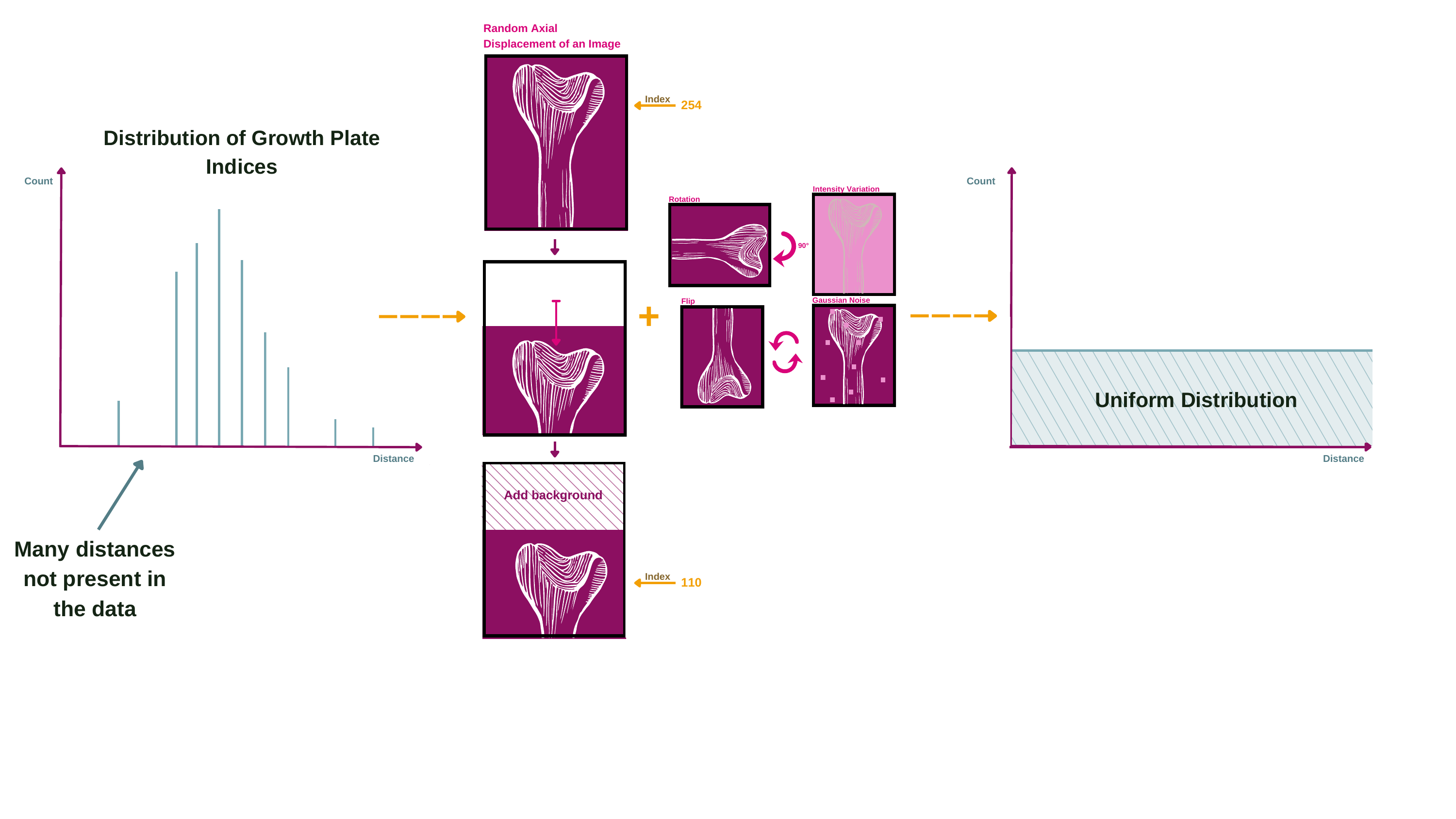}
  \caption{The data augmentation techniques used for solving the problem by MH team.}\label{fig:augmentation_matterhorn}
\end{figure}

To enhance model generalization, we implemented various data augmentation techniques, including $90$-degree rotations, horizontal flips, vertical flips, intensity variations, and Gaussian noise addition. More significantly, to achieve a uniform growth plane distribution across the entire domain, we applied axial plane displacement, filling with random background, a technique that proved vital for assuring proper generalization (see Figure~\ref{fig:augmentation_matterhorn} for more details).

Our final prediction was computed by averaging predictions from models trained in each cross-validation fold. For each of these individual models, we calculated several evaluations using different sets of random slices that were also averaged.

\begin{figure}[ht]
  \centering\includegraphics[width=0.47\textwidth]{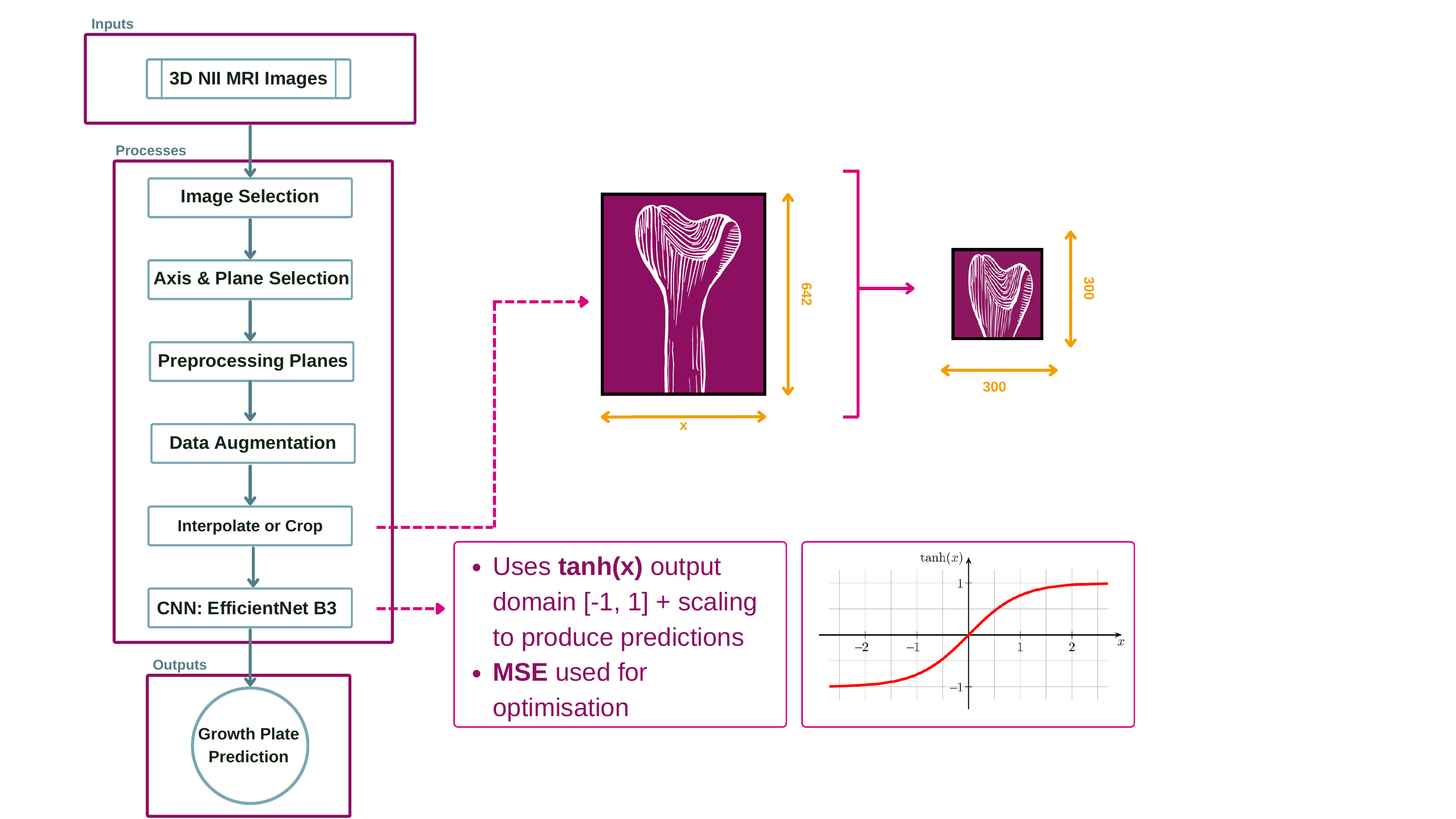}
  \caption{The whole training pipeline by MH team.}\label{fig:training_matterhorn}
\end{figure}

In Figure~\ref{fig:training_matterhorn} an illustration of the whole training pipeline is shown. This comprehensive methodology aimed to optimize model performance through hyperparameter tuning, pre-processing, data augmentation, stratified cross-validation and inference procedures.

\subsection{Team EK: Exploding Kittens}

\subsubsection{Image Preparation}

To predict the GPPI, we used a convolutional neural network (CNN) with a 2.5D approach. From each bone scan, $14$ different sagittal and coronal slices were taken, starting at the center $(x_0, y_0)$ and offsetting further slices by \(\pm 5\). Slices were then combined to create $3$-channel images, the first matching the sagittal slice, the second the coronal slice, and the third the blend of the two. This resulted in $49$ total images per scan to be used as input in a 2D CNN.

\smallskip
Input images were then resized to \(H \times W = 515 \times 515\) by cropping the diaphysis side and zero padding the other sides if necessary. A $\mu$CT bone window was applied by clipping HU to remove soft tissue and noise $W=2000, L=500$. Finally, pixel values were scaled and images saved as $8$-bit PNG using a naming convention to note image ID, offset from center, and GPPI.

\subsubsection{Model, Augmentation, and Parameters}

For our model we used an EfficientNet-es-pruned backbone, initialized with pretrained weights from the~\cite{deng2009imagenet} dataset. The head was set to a single output neuron resulting in a real number and a sigmoidal layer for $0-1$ scaling. To find the predicted growth plate plane, the model output is multiplied by the respective image size and rounded to the closest integer.

\begin{figure}
\centering
\includegraphics[width=0.7\linewidth]{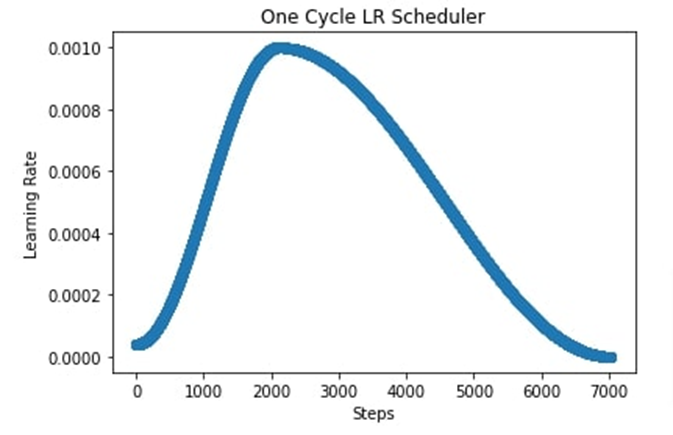}
\caption{The learning rate scheduler by EK team.}
\label{Figure 1. Learning rate scheduler overview.}
\end{figure}

\smallskip
Model training was done after unfreezing all layers with the following parameters:

\begin{itemize}
    \item Optimizer: AdamW
    \item Max learning rate: $1\times10^{-3}$
    \item Weight decay: $1\times10^{-2}$
    \item Learning rate scheduler: One cycle  (see Figure~\ref{Figure 1. Learning rate scheduler overview.})
    \item Loss: RMSE
    \item Epochs: $12$
    \item Batch size: $32$
\end{itemize}

\smallskip
When training the model, images were augmented to improve generalization by incorporating: RGB shift, random brightness and contrast, horizontal flip,  and normalization~\cite{deng2009imagenet}. This was done using Albumentations~\cite{info11020125}, an open source image augmentation library. Finally, to ensure proper sampling, four-fold cross-validation was run on the center slices of the images.

\begin{figure}
    \centering\includegraphics[width=1\linewidth]{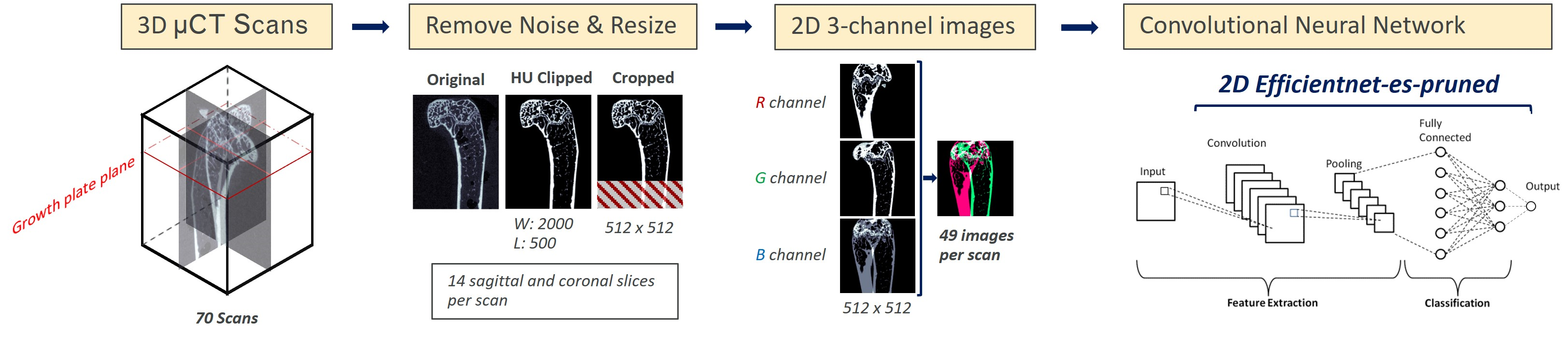}
    \caption{EK team's 2.5D approach.}
    \label{Figure 2. Exploding Kittens Model Summary}
\end{figure}

The overview of the training procedure is in Figure \ref{Figure 2. Exploding Kittens Model Summary}.

\subsection{Team CW: CodeWarriors2}

\begin{figure}
    \centering\includegraphics[width=1\linewidth]{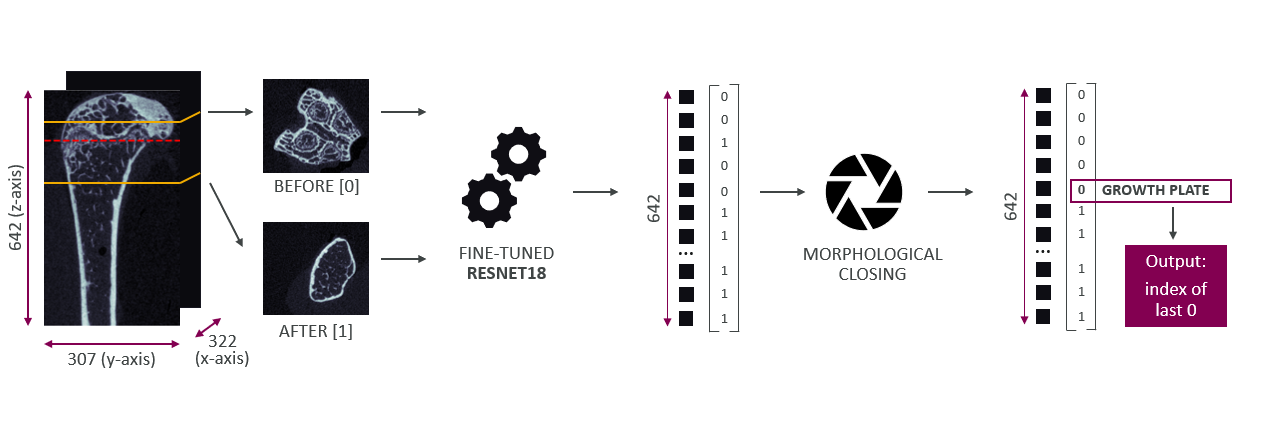}
    \caption{CW team's training process.}
    \label{fig:code-warriors-diagram}
\end{figure}

To detect GPPI of the bone, a pretrained deep learning model based on ResNet18 architecture~\cite{he2016deep} was utilized. The model was fine-tuned on the provided mice bone 3D imaging data. The process of determining GPPI is presented in Figure~\ref{fig:code-warriors-diagram}. The process consists of three steps: 

\begin{enumerate}
    \item Performing binary classification of all axial plane bone slices that were assigned to each image one of the two classes: before GPPI and after the GPPI.
    \item Applying morphological closing filters to denoise the sequence of outputs of the binary classifier to make the junction between the 'before' class and the 'after' class explicit.
    \item Assigning an index of the last image that is classified as 'before' as the GPPI. 
\end{enumerate}

Given that the train set images had different shapes, their size was reduced along $x$ and $y$ axes, this resizing not only ensured input consistency, but also lessen amount of background noise. To achieve this, each axial plane was masked using HU range corresponding to bone tissue as logical condition. Then normalized sum was calculated (the sum of $1$s corresponding to the bone values divided by the area) per each slice index for each $\mu$CT scan. First and last indexes for which this value was above threshold set to $0.03$ (meaning the bone values were on $3$\% of the slice area) were considered cutoff coordinates. 

ResNet18~\cite{he2016deep} model architecture was leveraged and fine-tuned for the provided data format and statement. The weights of the model were initialized from the ResNet18 model pretrained on the ImageNet dataset~\cite{deng2009imagenet}.  First convolutional layer was modified by replacing $3$ channel kernel with $1$ channel kernel in order to accept the grayscale input. Kernel's width and height remained the same.  The pretrained network classifies object into $1000$ categories, therefore, the last linear layer output dimension was replaced from $1000$ to $2$.

During training, image augmentations were applied to enhance better generalization by the model. The augmentations included random horizontal and vertical flips with probabilities of $0.5$ and random rotation up to $180$ degrees.

The model was trained using the AdamW~\cite{loshchilov2017decoupled} optimizer with a learning rate of $0.001$ and weight decay of $0.01$. Cross-entropy loss was utilized as a loss function and the batch size was set to $64$. 

To assess the performance of the model, five-fold cross-validation strategy was employed, where the number of training epochs was set to $15$ for all the folds. The final model was trained on the entire training set.

\subsection{Team SV: Subvisible}

The solution is based on the observation that prior to reaching the GPPI, a series of images in the axial plane present four distinct blobs, each corresponding to a protrusion. These blobs begin to merge upon reaching the GPPI. We trained a binary classifier to identify images with these four blobs. This classifier identifies a series of images preceding the GPPI as positive cases i.e. containing four blobs. The last index in this series serves as a rough estimate of the GPPI. 

To refine these initial estimates, we leveraged information on the axial axis and performed a search around these points. For this purpose, we employed a regression CNN. We examined $25$ slices before and after the initial rough estimates, resulting in a stack of $51$ images. This stack was fed into the model, the aim of which was to generate refined predictions of the GPPI. 

\begin{figure}[ht]
  \centering\includegraphics[width=0.47\textwidth]{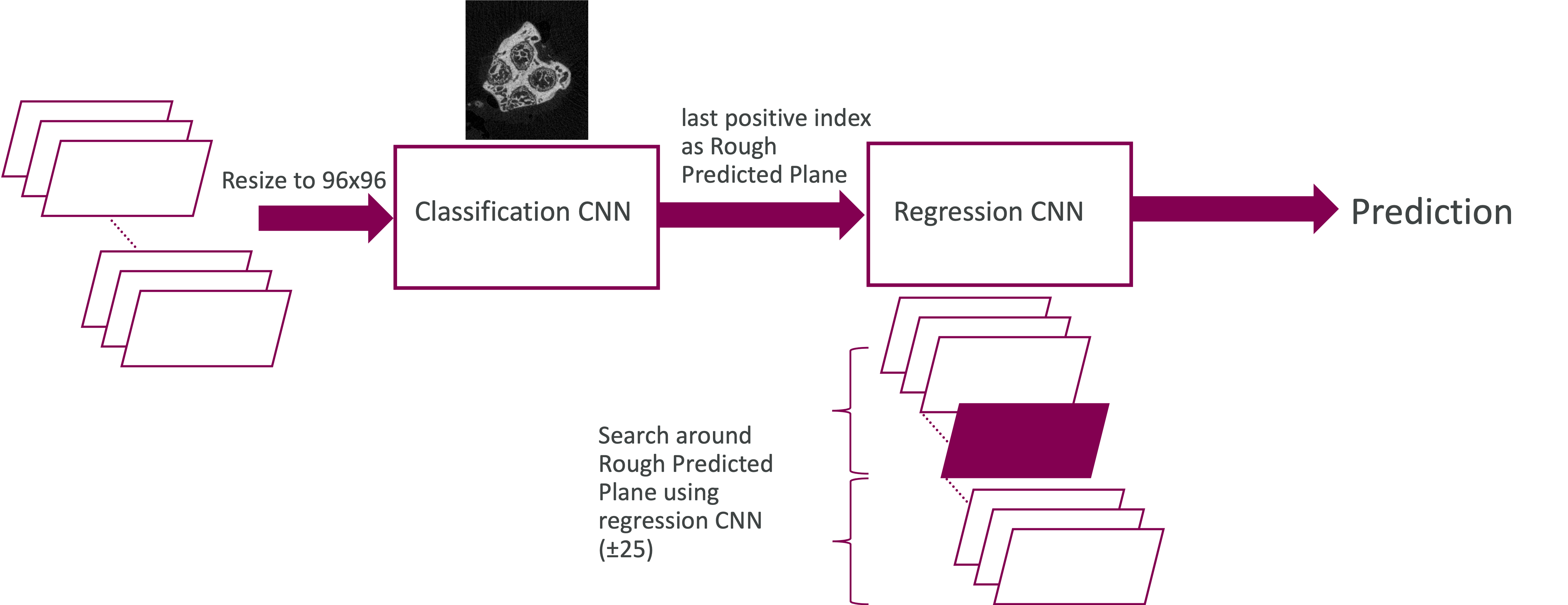}
  \caption{Diagram of the proposed approach by SV team.}\label{fig:method_subvisibe}
\end{figure}

\begin{figure}[ht]
  \centering\includegraphics[width=0.40\textwidth]{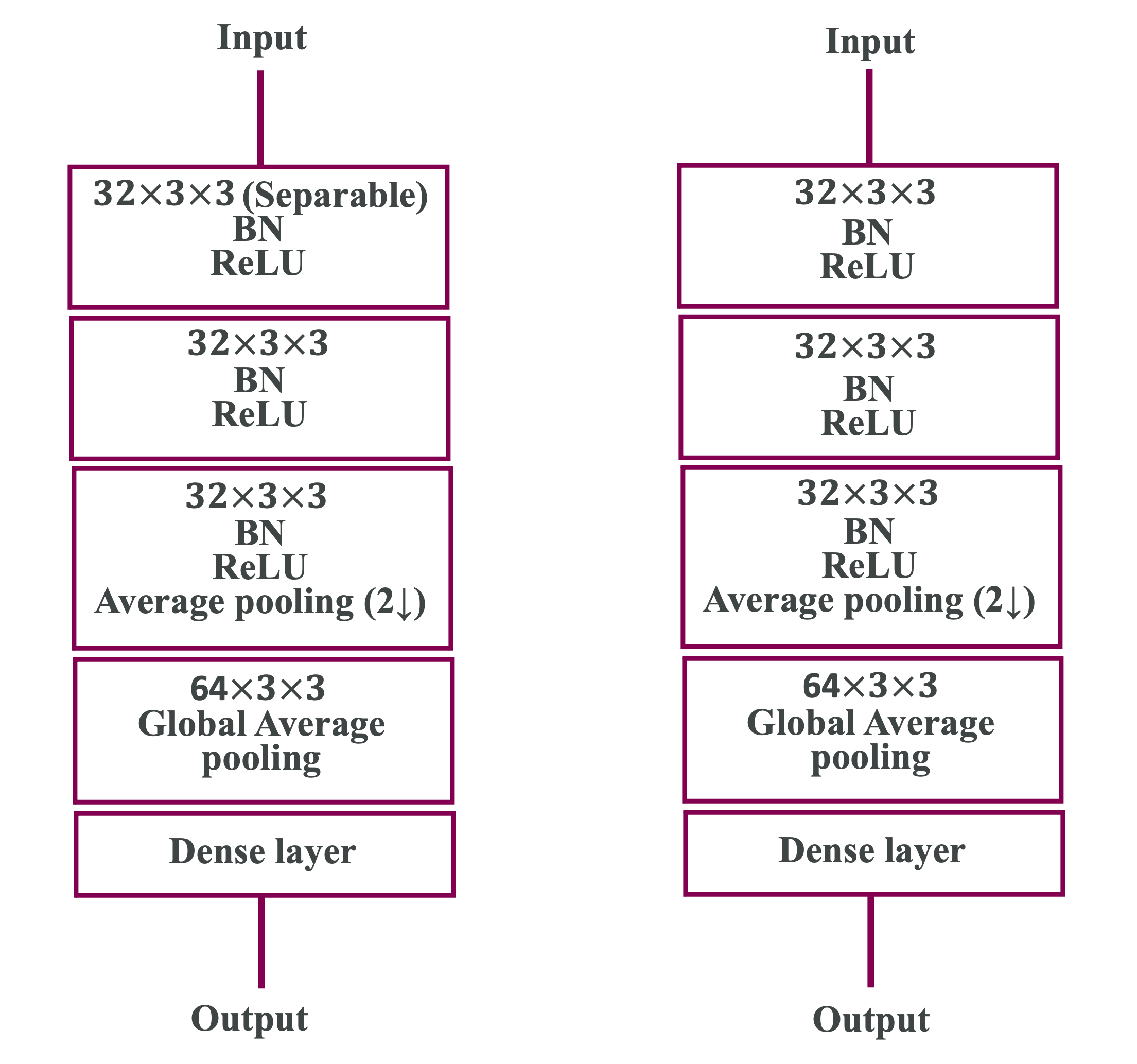}
  \caption{Network configuration for regression (left) and classification (right) by SV team.}\label{fig:config_subvisibe}
\end{figure}

Figure~\ref{fig:method_subvisibe} shows the overall diagram of our proposed approach. Considering the limited amount of training data, we trained two lightweight CNN networks for both the classification and regression tasks. Each network as shown in Figure~\ref {fig:config_subvisibe} has only four levels of convolution layers and less than $40$K parameters in total. These lightweight networks, along with resizing the images to $96\times96$ before feeding them to the models, make the training process quite fast. The loss function for the classification network is cross-entropy loss, and the negative of the score function is used as the loss function for the regression network. We incorporated data augmentation during training by randomly rotating images by $90$ degrees and flipping images (left/right). L2 regularization with a value of $0.0001$ was applied to the convolutional kernel during training. The models were trained using the Adam optimizer~\cite{kingma2015adam} for $120$ epochs, with a learning rate of $0.001$ for the first $80$ epochs and $0.0005$ ($0.0001$ for the classification network) for the remaining $40$ epochs. 

In generating the final predictions on the test set, a single binary classifier was used to create initial, rough estimates. However, to fine-tune these estimates, we leveraged an ensemble of five regression CNNs trained on the five different folds.

\subsection{Team BM: ByteMeIfYouCan}

\subsubsection{Training: Data Loader}

\begin{enumerate}
    \item We load the full 3D images, and as a first step scale the original HU values from the range $[-100, 3174]$ to the unit interval $[0, 1]$. This allows us to isolate the bone structure, and remove background noise.
    \item We ensure that all images have the same initial size, $(480\times480\times642)$, by either cropping or padding along the spatial dimensions. In all cases where cropping is necessary, only background is removed.
    \item We downsample the images by a factor of $2$ along the spatial dimensions, thus reducing them to a size of $(240\times240\times642)$. This allows for a more efficient the training process, while not sacrificing any relevant structural information. We do not perform any resizing along the depth dimension, along which we want to keep all available, small-scale information.
    \item We apply random cropping along the spatial dimensions to effectively create a shift augmentation, and apply a random crop to pick a $64$-plane window along the depth dimension. This reduces the image size to the final input size for training, $(224\times224\times64)$. We pick windows that contain or do not contain the annotated GPPI with equal frequency. A bigger window size means that more context information would be available to the model (e.g., parts of the GPP structure), but it would also be expected to generalize less well.
    \item We assign a target value $P=0\dots1$ for each window, according to the distance to the annotated GPPI, where $P=0$ if the GPPI is not contained in the window, $P=1$ if lies at the center, and $P$ varies linearly with distance from the center of the window to the GPPI.
    \item We apply some geometric augmentations along the spatial dimensions, including random flips and $90$ degree rotations, as well as random rotations in the range from $-10$ to $10$ degrees.
    \item We create batches of $32$ windows, containing $8$ random samples from $4$ images each, which are fed to the model during training and validation.
\end{enumerate}

\subsubsection{Training: Model, Loss \& Optimization}

We use a reduced DenseNet architecture as a regression model to predict target values $P$ for each input window, with $64$ initial features and a growth rate of $32$. We use $5$ stages with $4$ blocks each, to cover all spatial scales down to $7\times7$ pixels. In particular, we also use batch normalization and a dropout probability of $0.1$ for regularization. The resulting model has about $1.44$M trainiable parameters.

For the actual training, we randomly split the data into a training and a validation set, using 5 folds, i.e. a training fraction of $80$\%. We chose to calculate a BCE loss on $P$, and optimize it via Adam with a constant learning rate of $0.0001$. We initially trained for 500 epochs, but saw the validation loss stop decreasing significantly after epoch $200$, at which epoch we therefore continued to evaluate the models. We also calculate the evaluation scores on the validation sets across all folds, reaching an average score of about $0.548$ (see the main paper), and average offset between the actual and predicted GPPI of $-0.76$ pixels.

Due to time constraints arising from the timeline of the challenge, we did not perform an extensive hyper-parameter optimization, but rather made our choices based on intuition and some exploratory experimentation.

\subsubsection{Inference}

To actually predict a GPPI on unseen images, we use a sliding-window approach. We split the input image into windows, following steps $1-4$ from the training pipeline, except that in step $4$ we apply a center crop along the spatial dimensions instead of a random crop, and create windows shifted by $1$ px along the whole length of the bone. Having then predicted the target values $P$ for all of these windows, we can estimate a GPPI from the position of the window that yielded the maximum value.

For evaluation on the test set, we computed the rounded average GPPI from the values predicted by the individual models corresponding to each training fold, which yielded the final scores reported in the main paper.

\begin{figure}
    \centering\includegraphics[width=1\linewidth]{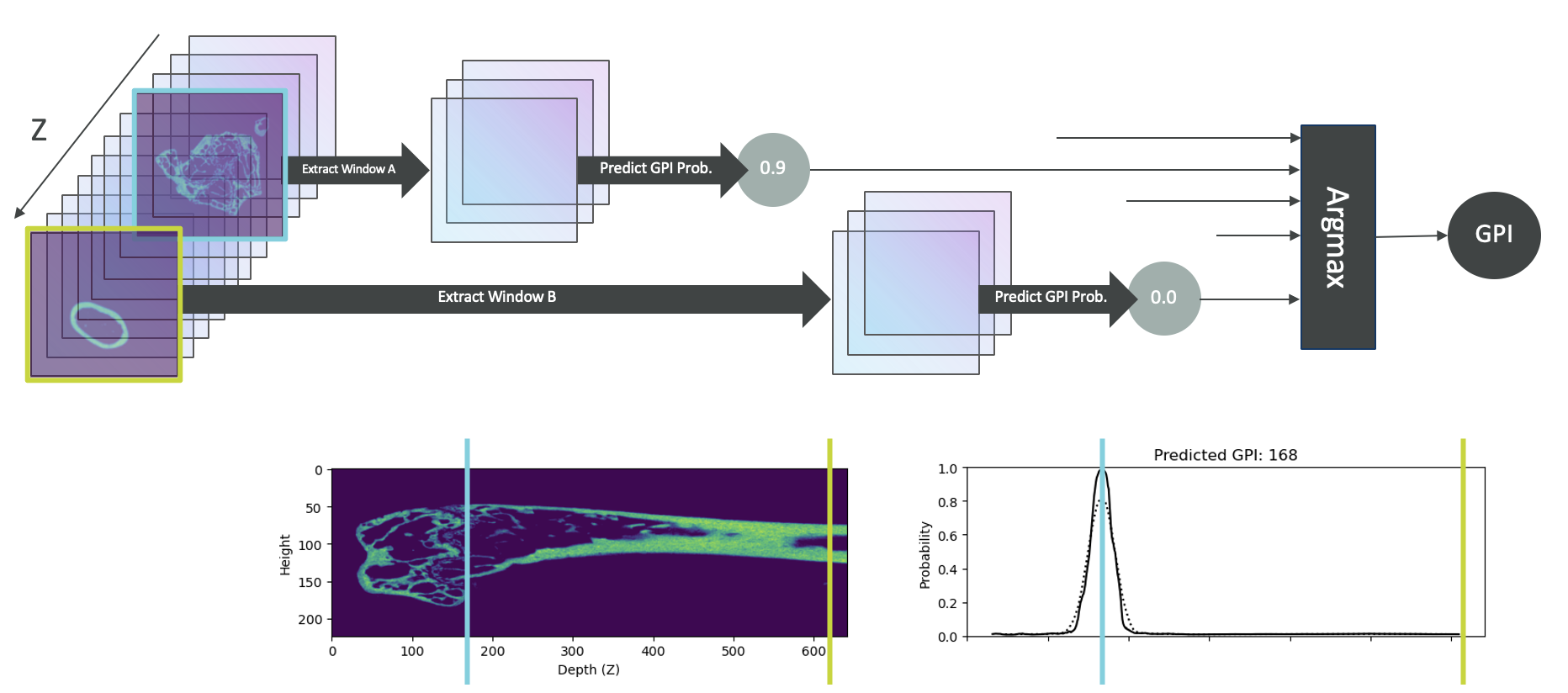}
    \caption{An illustration of the sliding window inference procedure by BM team.}
\end{figure}
\end{document}